\documentclass[preprint,12pt,fleqn]{elsarticle}

 \usepackage{graphics}

\usepackage[mathscr]{eucal}
\usepackage{amssymb}

\journal{Physica A}

\begin{document}

\begin{frontmatter}



\title{Heat transfer in rapidly solidifying
supercooled pure melt during final transient }


\author{G.L. Buchbinder \corref{cor1}}
\ead{glb@omsu.ru}
\author{V. A. Volkov }
\address{Department of Physics, Omsk State  University, Peace Avenue 55à, 644077 Omsk,
Russia }
 \cortext[cor1]{Corresponding author}

\begin{abstract}
The  heat transfer model  for a one-dimensional   supercooled melt
during the final stage  of solidification is considered.  The Stefan
problem  for the determination of the temperature distribution is
solved under the condition that (i) the interface approaches the
specimen surface with  a constant velocity $V$; (ii) the latent heat
of solidification  linearly depends on the interface temperature;
(iii) all the physical quantities given at the phase boundary are
presented by  linear  combinations of the exponential functions of
the interface position. First we find the solution of the
corresponding hyperbolic Stefan problem  within the framework of
which the heat transfer is described by the  telegraph equation. The
solution of the initial parabolic Stefan problem is then found as a
result  of the limiting transition $V/V_H \rightarrow 0$ $(V_H
\rightarrow \infty)$, where $ V_H $ is the velocity of the
propagation of the heat disturbances, in which the hyperbolic  heat
model teds to the parabolic one.

\end{abstract}


\end{frontmatter}


\section{Introduction}
\label{}

The process of rapid  solidification is a well established method
for the production of metastable solid states  of different nature
making it possible  to study  new mechanisms  of  crystal growth and
produce  materials  with radically new physical properties
\cite{HGH07}.

In  rapid solidification experiments very high velocities  of phase
interface  can be reached. Such conditions occur during
solidification  of the undercooled melts or recrystallization  after
pulsed-laser irradiation of a solid surface.  When the interface
velocity reaches  some critical value  diffusion-temperature field
in the bulk of  both phases can significantly deviate from local
equilibrium \cite{S95,GS97}.  In this case both the diffusion and
heat fluxes are no longer defined by the classical Fick's and
Fourier's laws  relating the diffusion and heat  fluxes
correspondingly to the gradients  of a solute concentration and
temperature.  The simplest generation of Fick's  and Fourier's laws
taking into account  the relaxation to local equilibrium in the
diffusion and the heat field is given by the Maxwell-Cattaneo model
and leads to the hyperbolic transport equations \cite{JC96}.

In the past two decades  a great body of studies, devoted to the
local nonequilibrium heat and mass transport during rapid
solidification, has been executed
\cite{S95,GS97,S96,SS95,S97,SS97,GD99,GD00,GPD001,GD002,G02,G04,GJ05,GH06,G07,GRH09}.
The numerical estimates show that under conditions of experimentally
achievable interface velocities local equilibrium is only disturbed
in the diffusion field, while the heat field can  be described in
the local-equilibrium approximation within the scope of the
conventional parabolic heat conduction model \cite{S95,GS97}

 The currently existing analytical models of the
directional solidification  processes usually consider  the initial
transient and the motion of the planar  front far from the
boundaries of a system \cite{D01,U85,W86}. The investigation of the
final transient is practically absent. Meanwhile, besides a purely
academic problem there exists  considerable practical interest as
well because the final study of solidification  process influences
the formation of the surface layer of the materials, their surface
physical-chemical characteristics and the distributions of different
defects {\cite{ZVS05,ZV06,HR10}.

The final transient  of a binary melt solidification  has been
analytically considered by Smith et. al. in the work \cite{STR55}
(also see  \cite{KF92}, p.278).  In the local equilibrium
approximation the authors have calculated the terminal solute
distribution of a formed solid. However the found distribution has
 divergence at the surface of the specimen  that may be caused,
among other reasons, by neglect of the temperature changes during
the interface motion  near the surface.  Because of the
thermodynamic relationship   between  the interface temperature and
the solute concentration, resulting from the phase diagram, such
changes can lead to further  solute redistribution in the bulk of
the specimen. In this connection it is of interest to initially
investigate the evolution of the temperature field of the
solidifying pure melt during the final transient.

In general, the problem of the determination of the temperature
field during  solidification of a melt  is known as the Stefan
problem. It consists in   solving of the heat conduction equation
for  a temperature $T$ of each phase with the boundary conditions at
the moving interface and in the one-dimensional form is written as
\begin{eqnarray}
&&\rho c_p\frac{\partial T_{L,S}}{\partial t} =
\lambda\frac{\partial^2T_{L,S}}{\partial x^2},\label{eq1}\\
&&T_L = T_S\,,\label{eq2}\\
&&q_L -  q_S = Q V\,,\label{eq3}
\end{eqnarray}
where indexes $L$ and $S$  are respectively related  to the liquid
and the solid
 phase, $\rho$ is the density, $c_p$  is the specific heat and $\lambda$ is  the thermal  conductivity
 (for simplicity, all  material characteristics  are assumed to be constant and identical
 within the phases and at the interface). Equation (\ref{eq2}) represents  the condition  of the continuity  of
the temperature  across the interface and equation (\ref{eq3})
defines the condition of a heat balance at the interface, where $
q_{L,S} = - \lambda \partial T_{L,S}/\partial x$ is the heat flux,
$Q$ is the latent heat of solidification  and $V$ is the velocity of
the interface. In addition,  at the surface of the specimen the
boundary conditions must be given.

 From  the thermodynamic point of view  the
  velocity  $V$  is determined by the undercooling of the
interface $\triangle T = T_m - T_i$, where $T_m$  is the equilibrium
temperature of solidification  and  $T_i$ is the interface
temperature, $V = f(\triangle T)$, so that  at $\triangle T = 0$, $V
= f(0) = 0$. For  so-called normal crystal growth it is assumed that
\begin{equation}
V = \mu\triangle T \label{eq4}
\end{equation}
where $\mu$  is  the kinetic coefficient characterizing  atomic
attachment kinetics at the interface \cite{D01}. When the
undercooling  of the interface $\triangle T $ is large enough, the
relationship (\ref{eq4}) can no longer hold. This takes place  when
the melt is initially supercooled. In a number of experiments with
pure metals, as well as by means of molecular-dynamic simulation
\cite{HB90,BH94,BH09,BGJ82} it has been shown that at the beginning
the growth velocity increases
 with increasing   undercooling $\triangle T $ reaching a maximum value and then
above some critical undercooling $\triangle T^* $, $V$  is
practically  kept constant in some region $\triangle T$ (also see
\cite{D01}, p.18). In what  follows  we shall presume  that  the
function $V = f(\Delta T)$ has   such properties.

When  the undercooling  is large enough the amount of  latent heat
released on the solidification front may prove to be deficient to
heat the interface  to the temperature $T_m$. In this case the
undercooling at the interface  will be different from zero during
the
 whole   solidification process, $\triangle T \neq 0$, and the
solid phase will reach the surface of the specimen with a finite
velocity, $V_f = f(\triangle T_f) \neq 0$. When the interface moves
in the near surface region the undercooling $\triangle T$, generally
speaking, will change, at the same time  also changing the growth
velocity $V$. However,  assuming  that $V_f$ is high enough, due to
  high undercooling \cite{HGH07}, and $\triangle T$
changes within the region for which velocity $V$ depends slightly on
$\triangle T$, one can consider that the  interface moves near the
surface with   approximately  constant velocity $V $  equal to
$V_f$. It should be noted that there exists a substantial
distinction from the situation  in the initial transient when the
interface velocity changes from zero to a steady state value.

The behavior of  the temperature $T_i$, the heat fluxes $q_{L,S}$
and the latent heat $Q$  at the moving interface  essentially
affects the evolution of the thermal field in the bulk of the
phases. Giving different models of their behavior at the interface,
one can consider various models of the solidification process. In
the present study we consider the exactly solvable model of
solidification within the scope of which any physical quantity $F$
(the temperature, the heat flux, etc.)  is given at the interface as
\begin{equation}
F(x, t)|_{t = t(x)}  =  A_0^{(F)} + A_1^{(F)}e^{- \gamma_1 x/2} +
A_2^{(F)}e^{- \gamma_2 x/2}\cdots\,, \label{eq5}
\end{equation}
where $t = t(x)$ determines  the path of the interface  position and
the coefficients $A_n^{(F)}$  and the powers  of the exponents
$\gamma_n$ must be defined  from the phase boundary conditions and
  the boundary condition at the surface of the system.   It is worth
  noting that an  expression of  similarly type has been obtained
  for the solute distribution  in the work \cite{STR55}.

As regards the latent heat  of the solidification  it is normally
assumed to be constant    and is empirically defined for the
equilibrium temperature $T_m$ as $Q = Q_m = kT_m $ $(k =
\textrm{const})$ (\cite{K76}, p.185). If the undercooling of the
interface is high enough,  it is reasonable to define the variable
latent heat as $Q = kT_i = Q_m - k\Delta T$  that  will be used in
what follows.

Thus  for the  determination  of the temperature field within the
scope of the given model we seek the solution of the one-dimensional
heat conduction equation (\ref{eq1}) with the boundary conditions
(\ref{eq2}), (\ref{eq3}) and (\ref{eq5}) when the interface
approaches  to the surface with a constant velocity $V$ and the
latent heat  linearly depends  on the undercooling $\Delta T$.

As  has been noted above, the temperature field in a rapid
solidifying
 melt can be considered within the scope of the parabolic model (\ref{eq1})-(\ref{eq3}).
However to obtain  the solution of interest  we shall initially
consider an auxiliary problem, namely,  the corresponding Stefan
problem for the hyperbolic heat conduction equation (see, for
example, \cite{SS95, GD002}). It turns out that the boundary
conditions given at the moving interface are more simply taken into
account within the scope of the hyperbolic model. As  is known, the
hyperbolic model of heat conduction, based on the telegraph
equation, gives the finite velocity of the propagation of the heat
disturbances in  matter $V_H$ and is reduced to the parabolic model
(\ref{eq1}) in the limit $V_H \rightarrow \infty $ \cite{JC96}. The
idea of the work is that initially the solution of the hyperbolic
Stefan problem with arbitrary ratio of the velocities $V/V_H < 1$ is
solved and then the limiting transient, $V/V_H \rightarrow 0$,  to
the solution of the parabolic problem is executed.

The work is organized as follows. In Sec.2  the hyperbolic  Stefan
problem corresponding to the  boundary problem
(\ref{eq1})-(\ref{eq3}), (\ref{eq5}) is considered. The solution of
the telegraph equation is found by the Riemann  method within the
scope of which the boundary conditions given at  arbitrary moving
boundary are automatically taken into account.  On the basis  of
this solution the heat fluxes and the temperature fields both in the
liquid phase and in the near interface region of the solid are
determined. The subsequent limiting  transition  $V/V_H \rightarrow
0$ and  the solution of the parabolic problem are given in Sec.3.
The conclusion is presented in  Sec.4. The Riemann method  and its
application to the presented  problem are contained in the
Appendices.

\section{\label{hyper}Hyperbolic model}
The hyperbolic model of the heat conduction starts  from the
Maxwell-Cattaneo relaxation equation for  the heat flux \cite{JC96}.
In the one-dimensional form this equation for the liquid phase is
\begin{equation}\label{eq6}
   q_L + \tau\frac{\partial q_L}{\partial t} = - \lambda \frac{\partial T_L}{\partial
   x}\,,
\end{equation}
where $\tau  = a/V_H^2$ is the time of the relaxation of the heat
flux to its local equilibrium value defined by the Fourier's law and
$a = \lambda /\rho c_p$  is the thermal diffusivity.

Equation (\ref{eq6}) in combination with the conservation law
\begin{equation}\label{eq7}
   \rho
c_p\frac{\partial T_L}{\partial t} = -  \frac{\partial q_L}{\partial
   x}\,,
\end{equation}
gives rise to the hyperbolic transport equations
\begin{eqnarray}
  \tau\frac{\partial^2 T_L}{\partial t^2} + \frac{\partial T_L}{\partial t} &=& a\frac{\partial^2 T_L}{\partial x^2} \label{eq8}\\
 \tau\frac{\partial^2 q_L}{\partial t^2} + \frac{\partial q_L}{\partial t} &=& a\frac{\partial^2 q_L}{\partial x^2}
 \label{eq9}\,.
\end{eqnarray}
The equation of the type (\ref{eq8}) or (\ref{eq9}) is known as the
telegraph equation. At $\tau \rightarrow 0$ (or $V_H \rightarrow
\infty $)
 the equations (\ref{eq8}) and (\ref{eq9}) are reduced to the parabolic heat conduction equation
  (\ref{eq1}).

Now let us consider the supercooled pure melt initially occupying
the half-space $x\geq 0$. The planar front of solidification forms
in the infinitely removed region at    $t = - \infty$
 and moves in parallel to the specimen surface fixed at
 $x = 0$. As it has been noted  in the Introduction when the undercooling
 is large enough the interface will move in the near surface region
 with the approximately  constant velocity
 $V = V_f$ along the path $x + Vt = 0$. In this case the region  occupied by the melt
 in the final stage of the solidification process is given by the
 inequality $0\leqslant x\leqslant - Vt$ $(t \leqslant 0)$. Therefore in the plane
 $(x, t)$ the liquid phase  occupies the region $x + Vt\leqslant0, x \geqslant 0, t
\leqslant 0$. At the interface  the condition of the heat balance
(\ref{eq3}) holds
\begin{equation}\label{eq10}
 (q_L  - q_S)|_{x + Vt = 0} = - VQ|_{x + Vt = 0}\,,
\end{equation}

Now we consider the heat flux $q_L$ in more detail. Introducing
dimensionless variables $t/\tau$, $x/\tau V_H$ in the equation
(\ref{eq9}), one obtains
\begin{equation}\label{eq11}
\frac{\partial^2 \tilde{q}_L}{\partial t^2} + \frac{\partial
\tilde{q}_L}{\partial t} = \frac{\partial^2 \tilde{q}_L}{\partial
x^2}\,,
\end{equation}
where   the former notations  $(x, t)$  have been used for new
variables, $\tilde{q}_L = q_L/(Q_mV_H)$  is a dimensionless heat
flux. The boundary condition (\ref{eq10}) in the dimensionless form
is written as
\begin{equation}\label{eq12}
(\tilde{q}_L  - \tilde{q}_S)|_{x + \alpha t = 0} = - \alpha
\tilde{Q}|_{x + \alpha t = 0}\,,
\end{equation}
where $\tilde{q}_S = q_S/(Q_mV_H)$, $\tilde{Q} = Q/Q_m$ and $\alpha
= V/V_H$ is the dimensionless parameter. In addition, we assume that
at the surface the equality  should be fulfilled
\begin{equation}\label{eq13}
   \tilde{q}_L(xt)|_{x=0} = 0\hspace{1cm}(t \leqslant 0)\,,
\end{equation}
expressing the condition of the absence  of the heat flux through
the surface. Finally, the solution of (\ref{eq11}) is sought in the
near surface region at $X \equiv x + \alpha t \leqslant 0$,
$x\geqslant 0$, $t\leqslant 0$ occupied by the liquid phase while
the solid occupies the region $X \geqslant 0$ (see fig. \ref{figA1}b
in Appendix).

Now we consider the case of $\alpha < 1$. Suppose that at the moving
interface residing in an arbitrary point $x$ near the surface at the
moment $t = - x/\alpha$ the flux $\tilde{q}_L$ and its  time
derivative $\partial \tilde{q}_L /\partial t$ are  known
\begin{equation}
 \tilde{q}_L(x t)|_{t = - x/\alpha}  = q_0(x)\hspace{0.5cm}
  \displaystyle\frac{\partial \tilde{q}_L(x t)}{\partial t}|_{t = - x/\alpha}  = q_1(x)\,,\label{eq14}
\end{equation}
where the functions  $q_0(x)$ and $q_1(x)$  will be specified
further.

If the functions $q_0(x)$ and $q_1(x)$ are known the solution of the
equation (\ref{eq11}) satisfying the conditions (\ref{eq14})  in the
region  $X \leqslant 0$ at $\alpha < 1$ can be found  by the Riemann
method \cite{TS04} (for details see Appendix A) and has the form
\begin{eqnarray}
&& \tilde{q}_L(xt)  =  \nonumber\\
&& = \frac{1}{2}\biggl \{\varphi \Bigl (- \alpha\frac{x + t}{1 -
\alpha}\Bigr)\exp\Bigl[\frac{X}{2(1 - \alpha)}\Bigr]  +
 \varphi \Bigl( \alpha\frac{x - t}{1 + \alpha}\Bigr)\exp \Bigl [-\frac{X}{2(1 + \alpha)}\Big]\biggr\} \nonumber \\
&&  - \frac{1}{2}e^{- t/2}\int\limits_{{-\frac{\alpha (x + t)}{1 -
\alpha}}}^{{\frac{\alpha (x - t)}{1 + \alpha}}}
  \,dx_1 \psi (x_1)e^{- x_1/2\alpha}J_0 \Bigl (\frac{1}{2}\sqrt{(x - x_1)^2 - (t + x_1/\alpha)^2}  \Bigr )\nonumber\\
 && + \frac{X}{4\alpha}e^{- t/2}\int\limits_{{-\frac{\alpha (x + t)}{1 - \alpha}}}^{{\frac{\alpha (x - t)}{1 + \alpha}}}
  \,dx_1 \varphi (x_1)e^{- x_1/2\alpha}\frac{J_0' \Bigl (\frac{1}{2}\sqrt{(x - x_1)^2 - (t + x_1/\alpha)^2}  \Bigr
  )}{\sqrt{(x - x_1)^2 - (t + x_1/\alpha)^2}}\,,\nonumber\\
  \label{eq15}
 \end{eqnarray}
where
\begin{equation}
  \varphi (x) = q_0(x)\,,\hspace{0.5cm}
  \psi (x) = \frac{1}{2}q_0(x) - \frac{1}{\alpha }q_0'(x) - \frac{1 - \alpha ^2}{\alpha
  ^2}q_1(x)\label{eq16}
\end{equation}
and $J_0(x)$ is the Bessel function of zero order.

In  accordance with what was said in the Introduction   all the
quantities given at the phase interface  are represented  by linear
combinations of the exponential functions (\ref{eq5}). In
particular, let $\varphi (x)$ and $\psi (x)$ be given  by the
expansions
 \begin{eqnarray}
  \varphi (x)  &=&  A_0 + A_1e^{- \gamma_1 x/2} + A_2e^{- \gamma_2 x/2}\cdots\,,\label{eq17}\\
 \psi(x) &=& B_0 + B_1e^{- \gamma_1 x/2} + B_2e^{- \gamma_2
 x/2}\cdots\,,\label{eq18}
 \end{eqnarray}
where constants $\gamma_n \geqslant 0$, $A_n$ and $B_n$ will be
specified in what follows. After the substitution of (\ref{eq17})
 and (\ref{eq18}) in  (\ref{eq15}) and the calculation of
 the integrals (details  in Appendix B), we obtain
\begin{equation}\label{eq19}
    \tilde{q}(x t) = \sum\limits_{n\geqslant 0}e^{- \gamma_n x/2}\Bigl\{ A_n^{(-)}\exp\Bigl[\frac{\gamma_n^{(+)}X}{2(1 - \alpha^2)}\Bigr ] +
    A_n^{(+)}\exp\Bigl[\frac{\gamma_n^{(-)}X}{2(1 - \alpha^2)}\Bigr ]
    \Bigr\},
\end{equation}
where the following notations have been introduced
 \begin{eqnarray}
   \gamma_n^{(\pm)} &=&  \gamma_n + \alpha \pm \sqrt{\alpha^2\gamma_n^2 + 2\alpha\gamma_n + \alpha^2} \geqslant 0\,; \label{eq20}\\
   A_n^{(\pm)}&=& \frac{A_n}{2} \pm  B_n\frac{\delta_n}{\nu_n} \,; \label{eq21} \\
   \delta_n &=& \frac{\alpha}{1 + \alpha\gamma_n}\,;\hspace{0.5cm} \nu_n=\sqrt{1 - \frac{\delta_n ^2}{\alpha^2}(1 - \alpha^2)}\,.\label{eq22}
\end{eqnarray}
Let us determine the parameters
 $\gamma_n$, $A_n$ and $B_n$  in such a way  as to
 satisfy the balance condition (\ref{eq12}) and the boundary
 condition at the sample surface (\ref{eq13}).

\subsection{\label{sec:par}The determination of the parameters }
Now consider the boundary condition (\ref{eq13}). Taking into
account that $\gamma _0 = 0$, $\delta_0 =\alpha$,  $\nu_0 = \alpha$,
$\gamma^{(\pm)}_0 = \alpha \pm\alpha$ and using the equation
(\ref{eq19}), we have for  arbitrary  small  $t < 0$
\begin{eqnarray}
\tilde{q}(x, t)|_{x = 0}&=&A_0^{(-)}\exp{\frac{2\alpha^2t}{2(1 -
\alpha^2)}} +  A_0^{(+)}\: + \nonumber\\
& +& A_1^{(-)}\exp{\frac{\gamma ^{(+)}_1\alpha t }{2(1 - \alpha^2)}}
+ A_1^{(+)}\exp{\frac{\gamma ^{(-)}_1\alpha t }{2(1 -
\alpha^2)}}\: + \nonumber\\
& + &A_2^{(-)}\exp{\frac{\gamma ^{(+)}_2\alpha t }{2(1 - \alpha^2)}}
+
   A_2^{(+)}\exp{\frac{\gamma ^{(-)}_2\alpha t }{2(1 - \alpha^2)}} + \cdots =
   0\,.\label{eq23}
\end{eqnarray}
If all the powers of the exponentials are different then
$\tilde{q}(0, t)= 0$ can be only at $A_n = B_n = 0$. However if each
exponential function  appears in the equation (\ref{eq23}) at least
twice then this can lead to nonzero $A_n$ and $B_n$. Bearing in mind
this circumstance we determine $\gamma_n$ so that the following
equalities  hold
\begin{equation}\label{eq24}
    \gamma^{(-)}_n = \gamma^{(+)}_{n - 1 }\hspace{2cm}n = 1, 2, 3,\ldots\;,
\end{equation}
in which $\gamma^{(+)}_{n - 1 }$ (and respectively $\gamma_{n - 1}$)
are considered to be known \footnote{The equation $\gamma^{(+)}_n =
\gamma^{(+)}_{n - 1 }$ either has no the solutions or does not give
the new ones.}. Taking into account  the notation (\ref{eq20}) and
resolving the equation (\ref{eq24}) in relation to $\gamma_n$, one
obtains
\begin{equation}\label{eq25}
    (1 - \alpha^2)(\gamma_n)_{12} = \gamma^{(+)}_{n - 1 } \pm \sqrt{\alpha\gamma^{(+)}_{n - 1 }[2(1 - \alpha^2) + \alpha\gamma^{(+)}_{n - 1
    }]\,}\:.
\end{equation}
At $n = 1$ and $ \gamma^{(+)}_0 = 2\alpha$  the equation
(\ref{eq25}) gives
\[\gamma_1 =  \frac{4\alpha}{1 - \alpha^2} .\]
The second value $\gamma_1 = 0$ is the extraneous root of the
equation (\ref{eq24}) at $n = 1$. After the determination of
$\gamma_1$ the values $\gamma^{(\pm)}_1$ appearing in (\ref{eq19})
can be found from the equation (\ref{eq20}). Along  a similar  line
one can obtain the values $\gamma_n$, $\gamma^{(\pm)}_n$ for $n
>1$. In
Table \ref{tab1} these values are given for $n \leq 4$. As is seen
from the table $\gamma_n \sim (1 - \alpha^2)^{- n}$,
 $\gamma_n^{(+)} \sim (1 - \alpha^2)^{- n}$ è $\gamma_n ^{(-)}\sim (1 - \alpha^2)^{- n +
 1}$.  The case of an arbitrary $n$ is easily  proved by induction
 using (\ref{eq25}).

\begin{table}
\centering
    \caption{The parameters of the equation (\ref{eq19})}
    \medskip
    \label{tab1}
\begin{tabular}{cccccccccc}
 \hline\hline
n&0&&1&&2&&3&&4\\
\hline
 $\gamma_n$ &$0$ & &${\frac{4\alpha}{1 - \alpha^2}}$ &&${\frac{4\alpha(3 + \alpha^2)}{(1 - \alpha^2)^2}}$
 &&${\frac{8\alpha (3 + \alpha^2)(1 + \alpha^2)}{(1 - \alpha^2)^3}}$
 &&${\frac{8\alpha(1 + \alpha^2)(\alpha^4 +10\alpha^2 + 5)}{(1 - \alpha^2)^4}}$ \\

 $\gamma_n^{(+)}$ &$2\alpha$   &&${\frac{8\alpha}{1 - \alpha^2}}$
 &&${\frac{2\alpha(3 + \alpha^2)^2}{(1 -
 \alpha^2)^2}}$&&${\frac{32\alpha(1 + \alpha^2)^2}{(1 - \alpha^2)^3}}$
 &&$  {\frac{2\alpha (\alpha^4 +10\alpha^2 + 5)^2}{(1 - \alpha^2)^4}}  $\\

$\gamma_n^{(-)}$&0  &&$2\alpha$ &&${\frac{8\alpha}{1 - \alpha^2}}$
&& ${\frac{2\alpha(3 + \alpha^2)^2}{(1 - \alpha^2)^2}}$
&&${\frac{32\alpha(1 + \alpha^2)^2}{(1 - \alpha^2)^3}}$\\
\hline
 \hline
\end{tabular}
\end{table}

\indent Under condition  (\ref{eq24}), the equation (\ref{eq23})
holds, if
\begin{eqnarray}
 A_0^{(+)}& = &A_0/2 + B_0  = 0\nonumber\\
  A_0^{(-)} &= &A_0/2 - B_0  = A_0\label{eq26}\\
A_n^{(+)}& =& -A_{n-1}^{(-)}\hspace{0.5cm} (n \geqslant 1)\nonumber
\end{eqnarray}
  Finally taking into account the equalities  (\ref{eq26}), the
  expression (\ref{eq19}) can be rewritten  in the form
\begin{equation}\label{eq28}
 \tilde{q}(x t) = \sum\limits_{n\geqslant 0}A_{n + 1}^{(+)}(e^{- \gamma_{n + 1} x/2} - e^{- \gamma_n  x/2})\exp\Bigl[\frac{\gamma_{n + 1}^{(-)}X}{2(1 - \alpha^2)}\Bigr ]
 .
\end{equation}

\subsection{\label{2.2} The temperature field}
The temperature field in the liquid phase can be found in the same
way as the heat flux has been defined.  The resulting expression for
the dimensionless temperature $\tilde{T}_L$ takes the form
\begin{equation}\label{eq31}
 \tilde{T}_L(x t) = a_0^{(+)} + \sum\limits_{n\geqslant 0}\{a_n^{(-)}e^{- \gamma_n x/2}
 +
 a_{n + 1}^{(+)}e^{- \gamma_{n + 1}  x/2}\}\exp\Bigl[\frac{\gamma_{n + 1}^{(-)}X}{2(1 - \alpha^2)}\Bigr
 ],
\end{equation}
where $\tilde{T}_L = \rho c_p(T_L - T_m)/Q_m$. The constants
$a_n^{(\pm)}$ can be expressed in  terms of the parameters
determining $T$ and $\partial T/\partial t$ at the interface by the
equations of the type (\ref{eq16})-(\ref{eq18}) and (\ref{eq21}).

Substituting the expressions for the flux (\ref{eq28}) and the
temperature (\ref{eq31}) into the energy consideration law
(\ref{eq7})  and equating  the coefficients at the linear
independent functions, one can express the constants $a_n^{(\pm)}$
in terms of $A_n^{(+)}$ appearing  in (\ref{eq28}). The
corresponding expressions will be given for the case of the
parabolic model.
\subsection{\label{2.3} The solid phase}
The heat flux $q_S$ and the temperature $T_S$ in the solid  satisfy
the equations
\begin{equation} \frac{\partial^2
\tilde{q}_S}{\partial t^2} + \frac{\partial \tilde{q}_S}{\partial
t}= \frac{\partial^2 \tilde{q}_S}{\partial
x^2};\hspace{1cm}\frac{\partial^2 \tilde{T}_S}{\partial t^2} +
\frac{\partial \tilde{T}_S}{\partial t}= \frac{\partial^2
\tilde{T}_S}{\partial x^2} \label{eq32}
\end{equation}
where  $\tilde{q}_S = q_S/(Q_mV_H)$,  $\tilde{T}_S = \rho c_p(T_S -
T_m)/Q_m$.

For the complete determination of the temperature  field in the
liquid the interface boundary conditions (\ref{eq2}) and
(\ref{eq12}) depending on the solid temperature and the heat flux
 must be used. For their
determination it will suffice to consider  the solutions of the
equations (\ref{eq32})  in the region near the interface defined by
the inequalities $X > 0$, $x + t < 0$  (see fig. \ref{figA1}c ). The
solutions of the equations (\ref{eq32}) in this region can be
obtained  in the same way  as  for  the liquid phase. The
application of the Riemann method  in the indicated region gives for
the heat flux
\begin{equation}\label{eq33}
    \tilde{q}_S(x t) = \sum\limits_{n\geqslant  0}e^{- \gamma_n x/2}\Bigl\{ \tilde{A}_n^{(-)}\exp\Bigl[\frac{\gamma_n^{(+)}X}{2(1 - \alpha^2)}\Bigr ] +
    \tilde{A}_n^{(+)}\exp\Bigl[\frac{\gamma_n^{(-)}X}{2(1 - \alpha^2)}\Bigr ]
    \Bigr\},
\end{equation}
where $\gamma_n^{(\pm)}$ are given by the equality  (\ref{eq20}) and
the constants $\tilde{A}_n^{(\pm)}$ can be expressed in terms  of
the parameters determining the  flux $\tilde{q}_S$ and its  time
derivative at the interface by the equalities of the type
(\ref{eq21}).

The expression for the temperature $\tilde{T}_S$ is analogously
written down as
\begin{equation}\label{eq34}
\tilde{T}_S(x t) = \sum\limits_{n\geqslant  0}e^{- \gamma_n
x/2}\Bigl\{ \tilde{a}_n^{(-)}\exp\Bigl[\frac{\gamma_n^{(+)}X}{2(1 -
\alpha^2)}\Bigr ] +
    \tilde{a}_n^{(+)}\exp\Bigl[\frac{\gamma_n^{(-)}X}{2(1 - \alpha^2)}\Bigr
    ]
    \Bigr\}.
\end{equation}
\section{The parabolic model}
As  has been  indicated above the transition  to the parabolic model
is executed by the limit $\alpha = V/V_H \rightarrow 0$ ($V_H
\rightarrow\infty $).  Using  table \ref{tab1}  and the equation
(\ref{eq25}) it is easy to show by the induction for any $n$ that
for  small $\alpha$
\begin{eqnarray}\label{eq35}
&& \gamma_n \simeq 2n(n + 1)\alpha\hspace{1cm}\alpha\rightarrow
0\qquad (\alpha \neq 0)\nonumber\\
&& \gamma_n^{(-)} \simeq 2n^2\alpha \\
&&\gamma_n^{(+)} \simeq 2(n + 1)^2\alpha \,.\nonumber
\end{eqnarray}
When the relationships (\ref{eq35}) are fulfilled the expressions
for the temperature $\tilde{T}_L$ and the flux $\tilde{q}_L$ in the
liquid phase are written down in the form
\begin{eqnarray}
&&\tilde{T}_L(x t) = a_0^{(+)} + \sum\limits_{n\geqslant 0}
\frac{A^{(+)}_{n + 1}}{\alpha (n + 1)} \{ e^{- \gamma_{n + 1}x/2 } +
 e^{- \gamma_n  x/2 } \} e^{ \gamma_{n + 1}^{(-)} X/2}\label{eq36}\\
 &&\tilde{q}_L(x t) =\sum\limits_{n\geqslant 0} A^{(+)}_{n
+ 1} \{ e^{- \gamma_{n + 1}x/2 }  -
 e^{- \gamma_n  x/2 } \} e^{ \gamma_{n + 1}^{(-)} X/2 }\label{eq37}
\end{eqnarray}
The constants $a^{(\pm)}_n$  in the equation (\ref{eq31}) for
$\tilde{T}_L$  have been defined  in such a way as to satisfy  the
conservation law (\ref{eq7}) (see the end of section 2.2).

From  expression (\ref{eq36})  it is seen that the disturbances of
the temperature field ahead of  the solidification front propagate
only over distances in the order of $l \lesssim 2\tau
  V_H/\gamma_1^{(-)} = a/V$  (in the dimensional variables).
  Therefore, if the interface  is removed from  the surface at the
  distance $l \sim a/V$, the surface still remains at the initial temperature $T_0$
\footnote{For example, for Ni   $a = 12\cdot 10^{- 6} m^2/s$ è $V
\sim 20 m/s$, $a/V \sim 0.6 \mu m$. \cite{BH94}.}. It is supposed,
of course, that  the constant interface velocity approximation holds
over distances  in the order of $a/V$ from the surface. At $x = 0$
and $V|t|\sim a/V$ in the expression (\ref{eq36}) one can neglect by
 sum ($|X| = |Vt| \sim a/V$) and  write down for the
temperature at the  surface
  \[\tilde{T}_L|_{x = 0} =  \Delta \thickapprox a_0^{(+)}  \: ,\]
where $\Delta = \rho c_p(T_0 - T_m)/Q_m < 0$ is  the initial
undercooling of the melt.

Now let us consider  the temperature field  in the solid phase. The
equalities (\ref{eq33}) and (\ref{eq34})  hold in the region between
the straight lines $x + \alpha t = 0$ and $x + t = 0$ (see fig.
\ref{figA1}c), or in the dimensional variables, between the straight
lines $x + Vt = 0$ è $x + V_Ht = 0$. At $V_H \rightarrow \infty$ the
second line goes  to the straight line $t = 0$ , $0 \leq x < \infty$
and the region of interest to us will be given by  the inequality $-
Vt < x < \infty$, spreading over the whole solid phase.

The variable part of the expressions  (\ref{eq33}) and (\ref{eq34})
in the dimensional coordinates  $(x, t)$ is determined by the
exponents
\[ e^{(n + 1)V[x + (n + 1)Vt]/a}\,, \hspace{0.5cm} e^{-nV[x - nVt]/a}\ \hspace{1cm}(V/V_H \ll 1)\, .\]
It is easy to see that at small $t$ the terms   containing the first
exponent (are proportional to $\tilde{A}_n^{(-)}$ or
$\tilde{a}_n^{(-)}$ ) with increasing  $x$  will indefinitely
increase. In order to avoid such nonphysical behavior we put
$\tilde{A}_n^{(-)} = \tilde{a}_n^{(-)} = 0$ and introduce the
notations $A_n^{(S)} = \tilde{A}_n^{(+)}$. Turning back to the
dimensionless coordinates, let us write down the expressions
(\ref{eq33}) and (\ref{eq34}) at  the small $\alpha$ in the form
\begin{eqnarray} &&\tilde{T}_S(x
t) = a^{(S)}_0 + \sum\limits_{n\geqslant 1}\frac{A_n^{(S)}}{\alpha
n}e^{- \gamma_n x/2}
    e^{\gamma_n^{(-)}X/2}\label{eq38}\\
  &&  \tilde{q}_S(x t) = \sum\limits_{n\geqslant 1}A_n^{(S)}e^{- \gamma_n x/2}
    e^{\gamma_n^{(-)}X/2},\label{eq39}
\end{eqnarray}
where the constants  $\tilde{a}_n^{(+)}$ have been determined so
that the conservation law (\ref{eq7}) is obeyed. It is easy to check
that the expressions (\ref{eq36})-(\ref{eq39}) satisfy the heat
conduction equations
\[
 \frac{\partial T_{LS}}{\partial t} = \frac{\partial ^2 T_{LS}}{\partial
 x^2}\,,\hspace{1cm}
 \frac{\partial q_{LS}}{\partial t} = \frac{\partial ^2 q_{LS}}{\partial x^2}
\]
and the Fourier's law is fulfilled, $\tilde{q}_{L,S} = -
\partial \tilde{T}_{L,S}/\partial x$.
\subsection{The temperature field}
For the determination parameters appearing in equations
(\ref{eq36})-(\ref{eq39}) we use  the condition of continuity  of
the temperature across the interface (\ref{eq2}). The detailed
calculations are given in Appendix C. The final expression for the
temperature of the liquid phase can be represented in the
dimensional coordinates $(x, t)$ as
\begin{eqnarray}\label{eq47}
 &&\tilde{T}_L(x, t) = \Delta + \frac{(1 + b\Delta)}{1 - b}\sum_{n \geqslant 1}C_n\Big \{e^{nV(x + nVt)/a} +  e^{- nV(x - nVt)/a}\Big
 \}\,,\\
 &&C_1 = 1\,; \hspace{0.5cm}C_n = b^{n - 1}\prod_{k = 2}^{n}\frac{1}{(2k - 1 - b)}\hspace{0.5cm} (n \geq 2);\nonumber\\
 &&( 0\leqslant x \leqslant -V t,\hspace{0.5cm} t \leqslant 0 )\,;\nonumber
\end{eqnarray}
where  $b = T_Q/T_m$ and $T_Q = Q_m/\rho c_p $. For metals the
dimensionless parameter $b$  varies through the range $0 < b < 1$.
For example, for Ni, $T_m = 1726 K, T_Q = Q_m/\rho c_p = 397K$ and
$b = T_Q/T_m = 0,23$ \cite{GD00}.

 Similarly one can write down for the solid phase
\begin{eqnarray}
&&\tilde{T}_S(x, t) = \frac{1 + \Delta }{1 - b} + \frac{1 + b\Delta
}{1 - b}\sum_{n \geqslant 1}\frac{2n + 1}{2n + 1 - b}C_n\,e^{- nV(x
- n Vt)/a}
\,,\label{eq48}\\
&&( - V t\leqslant x  ,\hspace{0.5cm} t \leqslant 0 )\,.\nonumber
\end{eqnarray}

 It is easily seen that each term in the
brace (\ref{eq47}) represents the superposition of two heat waves
propagating in the mutually opposing directions with the velocity
$nV$.
\subsection{Numerical results}
\begin{figure}
    \centering
    \includegraphics[width=2.6in]{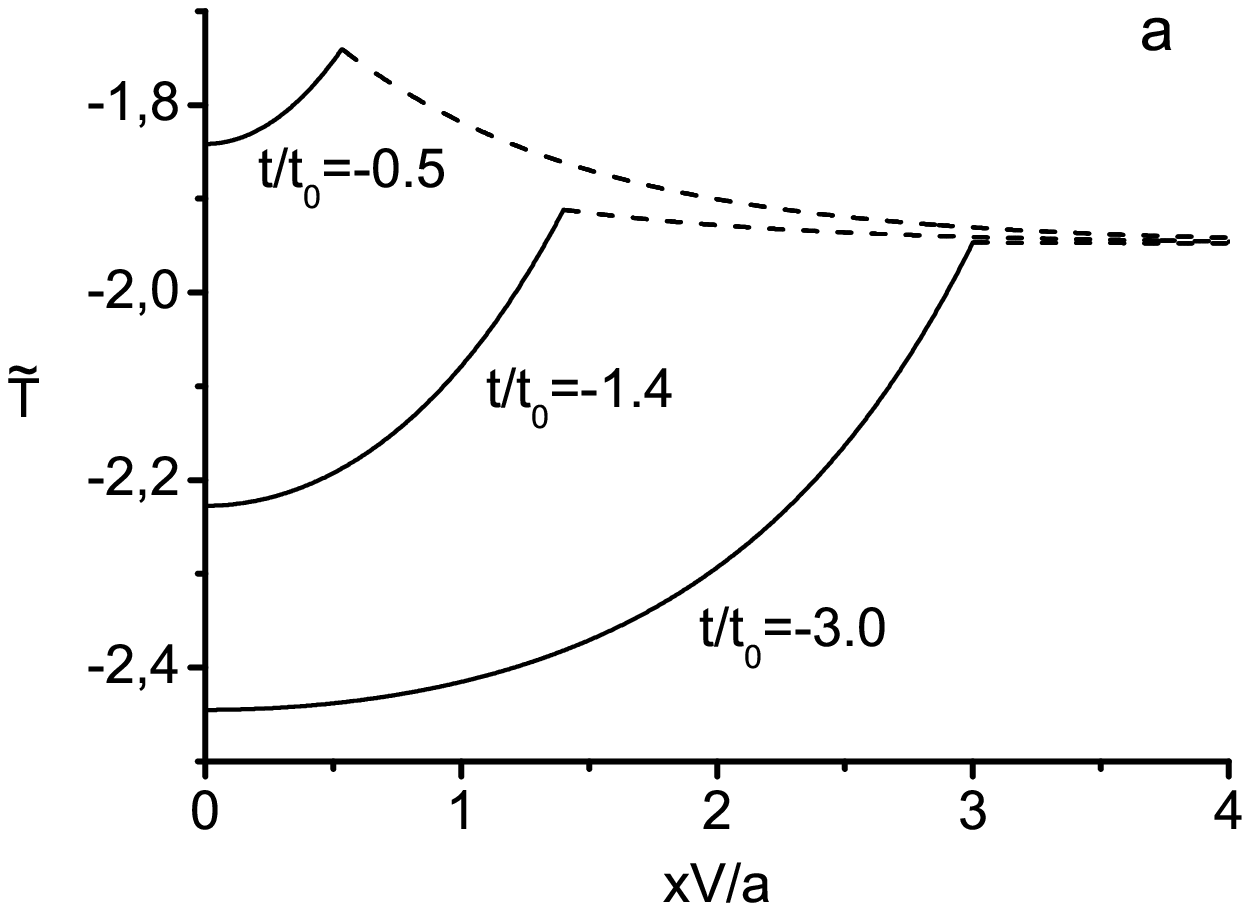}
    \includegraphics[width=2.6in]{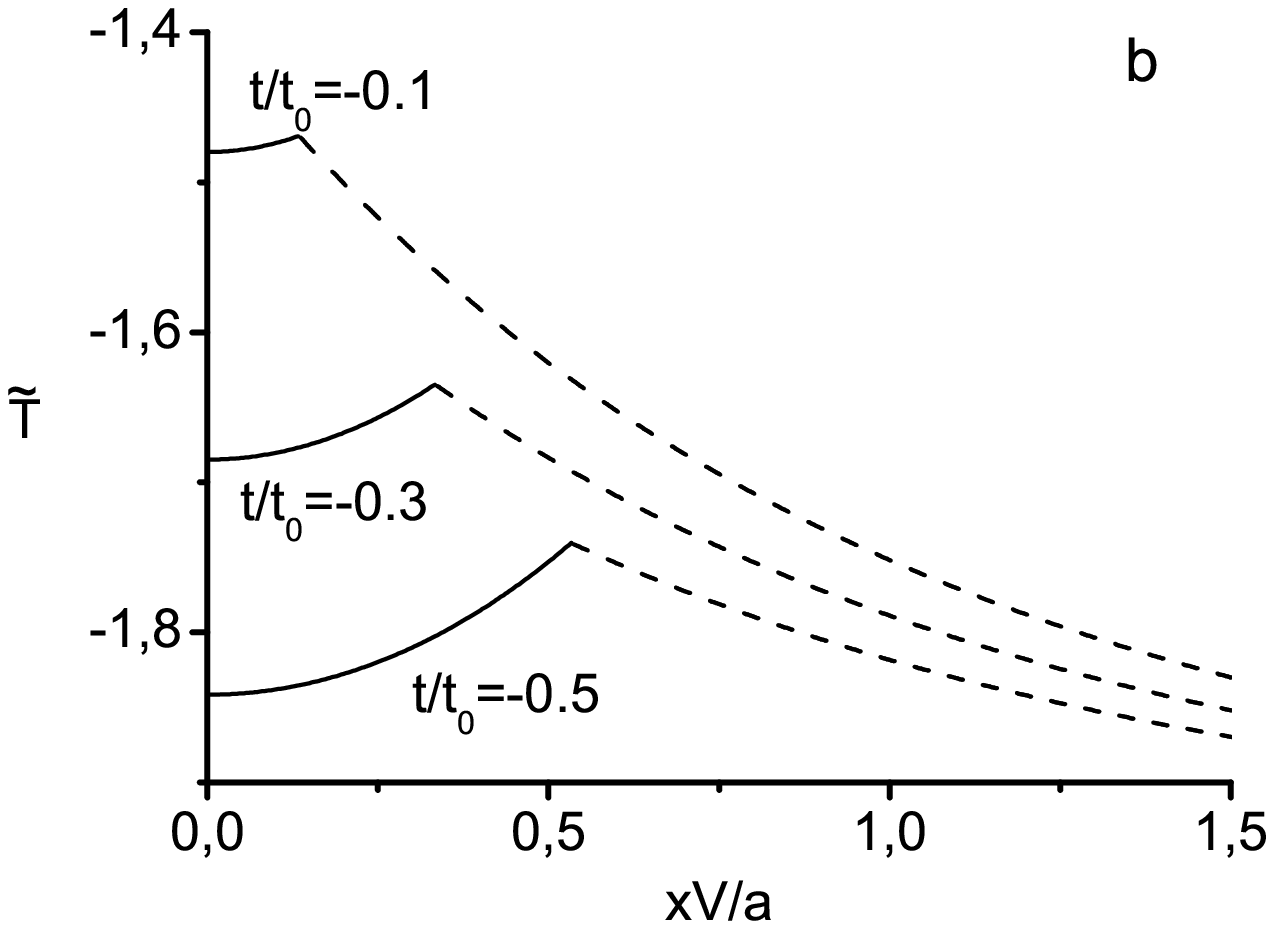}
    \caption{The temperature profiles for different moments of time. The dashed lines correspond to the
    temperature
     distribution in the solid phase, the solid lines correspond to the liquid phase; $t_0 = a/V^2$. }
    \label{fig1}
\end{figure}
Figures \ref{fig1}-\ref{fig2} present the temperature profiles
obtained from the equations (\ref{eq47}), (\ref{eq48}) for $\Delta =
- 2.5$ (supercooling) and $b =  0,23$.

The temperature curves for some moments of time are shown in figure
\ref{fig1}. The dashed lines are  the temperature distributions in
the solid phase, the solid lines give the temperature field in the
liquid phase. As  is seen from figure \ref{fig1}a when the interface
is relatively far  from the  surface  ($t/t_0 =- 3, t_0 = a/V^2$)
the temperature of the solid phase is constant, and the
 temperature of the liquid phase falls to  approximately the initial temperature of the melt $T_0$
($\tilde{T}_L|_{x=0} \thickapprox\Delta = (T_0 - T_m)/T_Q = - 2.5$)
at the surface. When the interface moves  close enough to thermal
isolated surface (figure \ref{fig1}b), the released latent heat
gives rise to the gradual heating of both the liquid phase and the
near-interface region of the solid.

In the figure \ref{fig2} the dependence  of the interface
temperature on the parameter b is shown.  From  the figure it is
also seen that the interface appears  on the surface (at $x = 0$) in
the supercooled state, $|T_i - T_m|/T_Q > 1$, providing  high final
velocity of the solidification processes $V = V_f$.
\begin{figure}
   \centering
   \includegraphics[width=3.5in]{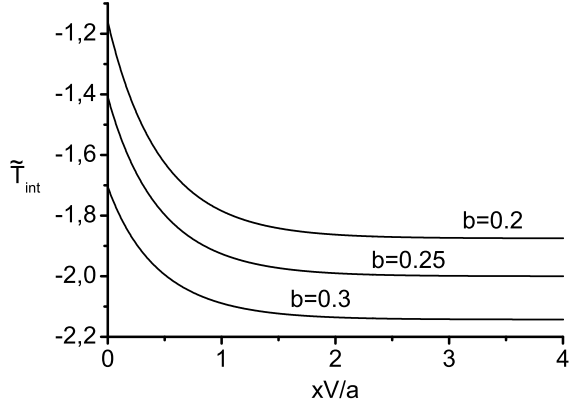}
    \caption{The temperature  at the interface depending on the
    distance to
   the free surface for different values of the parameter $b = T_Q/T_m$ .}
   \label{fig2}
\end{figure}
\section{Conclusion}
In the given work  we have considered a one-dimensional  model of
the heat conduction in the supercooled melt during the final
transient. Three main assumptions underlie   the model. Firstly, it
is supposed that  the interface  approaches  the system surface with
a constant velocity.  Some reasons for this assumption are provided
by a number of  experimental results and molecular-dynamic
simulation \cite{HB90,BH94,BH09,BGJ82,D01} showing that when the
undercooling of the melt is large enough  the interface velocity can
slightly depend on the temperature. The second supposition assumes
that the latent heat of  solidification linearly depends on the
interface temperature. Finally, it is supposed  that the physical
quantities of interest (the temperature, the heat flux, etc.)  given
at  the interface are presented by  linear combination of the
exponential functions of the form (\ref{eq5}), the parameters of
which are determined   as part  of the general solution of the
problem.

Within the scope  of the model  the exact  solution of the
one-dimensional Stefan problem (\ref{eq1})-(\ref{eq3}), (\ref{eq5})
defining  thermal distribution in the system when  the interface
moves near the surface  has been found.  To this end, initially, the
corresponding hyperbolic  Stefan problem  has been   considered
within the framework of  which the  heat transfer is described on
the basis of the telegraph equation. The telegraph equation for  the
heat flux and the temperature  in both  the liquid  phase  and the
near interface region  of the solid has been resolved  by the
Riemann method. Further  we have used  the fact that  in the limit
$\alpha = V/V_H \rightarrow 0$ the hyperbolic heat model is reduced
to the parabolic one. Taking into account  this circumstance and
executing the limiting transition $\alpha  \rightarrow
 0$ in the expressions  for the fluxes (\ref{eq28}), (\ref{eq33})
 and  the temperature (\ref{eq31}),
 (\ref{eq34}) the thermal distribution in the sample during the final stage of
 solidification
has been obtained.

In conclusion it should be noted, that the given approach allow us
to consider also other models of the solidification process
differing from  model (5). It is likely that the interface boundary
conditions in the form of the superposition of exponential functions
are the only ones for which  the exact solution exists. On the other
hand the solution for the heat flux (15) is written down for
arbitrary
  boundary conditions (arbitrary $\varphi$ and $\psi$) and
opens up the possibility of numerical simulation.

\clearpage
 \appendix
\section{The Riemann method}
\setcounter{figure}{0}
\begin{figure}
    \centering
    \includegraphics[width=5.5in ]{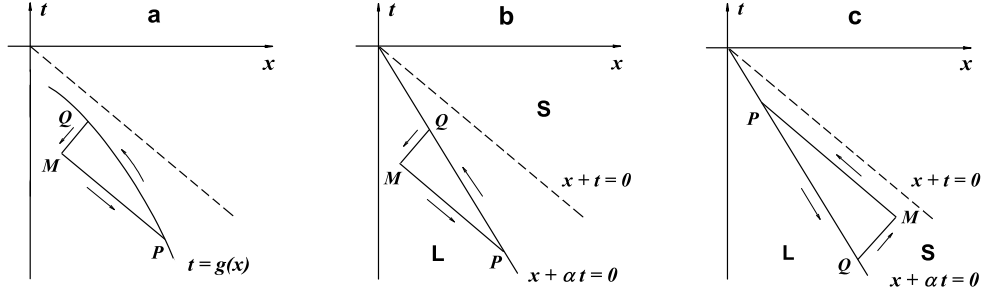}
    \caption{(a)The figure to the Riemann method. (b) The solution of the equation (\ref{eq11})  is sought
    in the region of the liquid phase $x + \alpha t < 0$.
    (c) The solutions of the equations (\ref{eq32}) are sought  in the region of the solid phase
   situated  between straight lines
    $X = x + \alpha t = 0$ and $x + t = 0$  .}
    \label{figA1}
\end{figure}
Let it be required to find the solution of the linear hyperbolic
equation
\begin{equation}\label{eqA1}
\frac{\partial ^2 \tilde{q}}{\partial t^2} + \frac{\partial
\tilde{q}}{\partial t} = \frac{\partial ^2 \tilde{q}}{\partial
x^2}\,,
\end{equation}
satisfying  the initial conditions given at the curve
 $\Gamma$ :
$t = g(x)$ (see figure \ref{figA1}a)
\begin{eqnarray*}
\tilde{q}|_{t = g(x)} &=& q_0(x) \\
 \frac{\partial \tilde{q}}{\partial t}\Bigr |_{t = g(x)} &= &q_1(x)\,.
\end{eqnarray*}
The substitution
  $\tilde{q} = e^{- t/2}u$
makes it possible to lead  equation (\ref{eqA1}) to a more simple
form
\begin{equation}\label{eqA2}
\frac{\partial ^2 u}{\partial x^2}  - \frac{\partial^2 u}{\partial
t^2}  + \frac{1}{4}u = 0\,,
\end{equation}
with the initial conditions
\begin{eqnarray}
  u|_{t = g(x)} &=& q_0(x)e^{g(x)/2} \equiv \varphi_1 (x)\label{eqA3} \\
  \frac{\partial u}{\partial t}\Bigr |_{t = g(x)} &=& ({\textstyle\frac{1}{2}}q_0 +q_1)e^{g(x)/2} \equiv \psi_1(x)\,.\label{eqA4}
\end{eqnarray}
The characteristics of equation (\ref{eqA2}) are the straight lines
$x \pm t = \mbox{const}$. According to the Riemann method
\cite{TS04} if the characteristics go through the point $M$ and
intersect with the curve $\Gamma$ at the points $P$ and $Q$ , then
the solution of equation (\ref{eqA2}) at the point $M$ can be
represented as

\begin{eqnarray}
  u(M) &=& \frac{1}{2}\bigr (u_P + u_Q  \bigl) - \nonumber\\
  &- & \frac{1}{2}\int\limits_{PQ}v\Bigl ( \frac{\partial u}{\partial x_1 }dt_1 + \frac{\partial u}{\partial t_1} dx_1 \Bigr
   ) - u\Bigl ( \frac{\partial v}{\partial x_1 }dt_1 + \frac{\partial v}{\partial t_1} dx_1 \Bigr
   )\label{eqA5}
\end{eqnarray}

The integral in (\ref{eqA5}) is taken along the curve  $\Gamma$ from
 $P$ up $Q$ and $u_P$ and $u_Q$ are the values of  $u$, taken at the points  $P$
 and $Q$. The Riemann function $v(M, M_1)$ for equation(\ref{eqA3}) has the  form
\begin{equation}\label{eqA6}
    v(M, M_1) = J_0\Bigl ({\textstyle\frac{1}{2}}\sqrt{(x - x_1)^2 - (t -
    t_1)^2}
    \Bigr )\,,
\end{equation}
where $J_0(x)$  is the Bessel function of zero order and ${\partial
u}/{\partial x}$ is calculated along the curve as
\begin{equation}\label{eqA7}
    \frac{\partial u}{\partial x}\Bigr |_{t = g(x)} = \varphi
    _1'(x) - \psi_1(x)g'(x)\,.
\end{equation}
The Riemann method for arbitrary linear hyperbolic equations  can be
found, for example, in \cite{TS04}.

Now consider the solution of  equation (\ref{eqA3}) in the region $x
\geqslant 0$, $t \leqslant 0$, $X = x + \alpha t < 0 $, when  the
initial data are given at the straight line  $ t = - x/\alpha$ (see
Figure \ref{figA1}b). Instead of (\ref{eqA3}) and (\ref{eqA4}) we
have
\begin{eqnarray}
 u|_{t = - x/\alpha} &=& q_0(x)e^{-x/2\alpha} \label{eqA8} \\
  \frac{\partial u}{\partial t}\Bigr |_{t = - x/\alpha} &=& ({\textstyle\frac{1}{2}}q_0 +q_1)e^{-x/2\alpha} \,.\label{eqA9}
\end{eqnarray}
If the point $M$ has coordinates  $(x, t)$, so it is easy  to show
that the points $P$ and $Q$ have the abscissas respectively equal to
\begin{equation}\label{eqA10}
    x_P = - \frac{\alpha (x + t)}{1 - \alpha}\, ;\hspace{1 cm} x_Q =  \frac{\alpha (x - t)}{1  + \alpha }
\end{equation}
Consider the integral term in equation (\ref{eqA5}). Using equations
(\ref{eqA7})-(\ref{eqA10}) and the fact that along the pathway of
integration $dt_1 = - dx_1/\alpha$, one has

\begin{equation}\label{eqA11}
\frac{1}{2}\int\limits_{-\frac{\alpha (x + t)}{1 -
\alpha}}^{\frac{\alpha (x - t)}{1 + \alpha}}dx_1 e^{-
x_1/2\alpha}\Bigl \{v\psi (x_1) + \varphi  (x_1) \Bigl (
\frac{1}{\alpha }\frac{\partial v}{\partial x_1} - \frac{\partial
v}{\partial t_1} \Bigr ) \Bigr\}_{t_1 =- x_1/\alpha}\,,
\end{equation}

where the notations are introduced
\begin{eqnarray}
  \varphi (x) &=& q_0(x)\,, \nonumber \\
  \psi (x) &=& \frac{1}{2}q_0(x) - \frac{1}{\alpha }q_0'(x) - \frac{1 - \alpha ^2}{\alpha
  ^2}q_1(x)\,.\nonumber
\end{eqnarray}
Furthermore using  the Riemann function (\ref{eqA6}), it can show
that
\begin{equation}\label{eqA12}
\Bigl (\frac{1}{\alpha }\frac{\partial v}{\partial x_1} -
\frac{\partial v}{\partial t_1}\Bigr )\Bigl |_{t_1 =
 - x_1/\alpha} =
 - \frac{X}{2\alpha} \frac{J_0' \Bigl
(\frac{1}{2}\sqrt{(x - x_1)^2 - (t + x_1/\alpha)^2} \Bigr
  )}{\sqrt{(x - x_1)^2 - (t + x_1/\alpha)^2}}.
\end{equation}
Finally, after  substitution of integral (\ref{eqA11}) into equation
(\ref{eqA5}) and using  the equality $\tilde{q} = e^{- t/2}u$, one
obtains the solution of the starting equation (\ref{eqA1}), with
added conditions (\ref{eq14}), in the form represented by
(\ref{eq15}).

\section{The calculation of the integrals}

Substituting equations (\ref{eq17}) and (\ref{eq18}) into
(\ref{eq15}) we have
\begin{equation}\label{eqB1}
\tilde{q}_L(x, t) = \sum\limits_{n\geqslant 0}\tilde{J}_n(x, t)\,,
\end{equation}
where

\begin{eqnarray}
& &\tilde{J}_n(x, t)  = -  B_nJ_n^{(1)} +A_nJ_n^{(2)} + \nonumber\\
 &&\phantom{aa} + \frac{A_n}{2}\Bigl\{\exp\Bigl[\frac{\alpha\gamma_n(x + t) +
X}{2(1 - \alpha)}\Bigr ]  + \exp\Bigl[- \frac{\alpha\gamma_n(x - t)
+ X}{2(1 + \alpha)}\Bigr ] \Bigr\}\,\label{eqB2}
\end{eqnarray}
and
\begin{eqnarray}
  J_n^{(1)} &=&  \frac{1}{2}e^{- t/2}\int\limits_{-\frac{\alpha (x + t)}{1 - \alpha}}^{\frac{\alpha (x - t)}{1 + \alpha}}
  dx_1 e^{- x_1/2\delta_n}J_0 \Bigl (\frac{1}{2}\sqrt{(x - x_1)^2 - (t + x_1/\alpha)^2}  \Bigr )\,;\nonumber\\
  \label{eqB3}\\
J_n^{(2)}&=& \frac{X}{4\alpha}e^{- t/2}\int\limits_{-\frac{\alpha (x
+ t)}{1 - \alpha}}^{\frac{\alpha (x - t)}{1 + \alpha}}
  dx_1 e^{- x_1/2\delta_n}\frac{J_0' \Bigl (\frac{1}{2}\sqrt{(x - x_1)^2 - (t + x_1/\alpha)^2}  \Bigr
  )}{\sqrt{(x - x_1)^2 - (t + x_1/\alpha)^2}}\,;\nonumber\\ \label{eqB4}\\
 \delta_n &=& \frac{\alpha}{1 + \alpha\gamma_n}\,.\label{eqB5}
\end{eqnarray}

\subsection*{The calculation of $J^{(1)}_n$}
Making the substitution in the integral (\ref{eqB3})
\[\frac{2\alpha X}{1 - \alpha ^2}z = \frac{\alpha (x + t)}{1 - \alpha} +
x_1\,,\] we have (for convenience the index $n$ is omitted)
\begin{equation}\label{eqB6}
    J^{(1)} = \frac{\alpha X}{1 - \alpha ^2}\exp \Bigl [ \frac{X'}{2(1 - \alpha
    )}\Bigl ]\mathscr J\,,
\end{equation}
where the following notations are introduced
\begin{eqnarray}
  &{\mathscr J} &= \int\limits ^1_0 e^{- \mu z} J_0\Bigl (\beta \sqrt{z(1 - z)}\Bigr )dz \,,\label{eqB7} \\
  &X'&= X + \Bigl (\frac{\alpha}{\delta} - 1 \Bigr )(x + t)\label{eqB8}\, ,  \\
 \mu &=& \frac{\alpha X}{\delta (1 - \alpha ^2)} < 0\,,\quad  \beta = - \frac{X}{\sqrt{1 - \alpha ^2}} > 0\,. \label{eqB9}
\end{eqnarray}
Consider the integral ${\mathscr J}$. Using the definition of the
Bessel function
\[ J_0 \Bigl ( \beta\sqrt{z - z^2}\Bigr ) = \sum\limits^\infty _{m = 0}\frac{(-1)^m (\beta /2)^{2m}(z - z^2)^m }{m!\,\Gamma (m + 1)} \,,\]
where $\Gamma (x)$  is the Euler gamma-function, one represents  the
integral (\ref{eqB7}) in the form
\begin{equation}\label{eqB10}
{\mathscr J} = \sum\limits ^\infty _{m = 0}\frac{(-1)^m (\beta
/2)^{2m}}{m!\,\Gamma (m + 1)}\int \limits _0^1 e^{- \mu z}(z -
z^2)^mdz
 \end{equation}
Calculating  the latter integral \cite{GR71}, one obtains
 \begin{equation}\label{eqB11}
{\mathscr J} =  \Bigl ({\pi}/{|\mu|}\Bigr )^{1/2}\; e^{-
\mu}\sum\limits_{n = 0}^{\infty}\frac{(- \beta ^2/4|\mu|)^m}{m!}I_{m
+ 1/2}\Bigl (\frac{|\mu|}{2} \Bigr )\,,
 \end{equation}
where $I_{\nu}(x)$ is the modified Bessel function of the first
kind.  Furthermore, we use the equality \cite{PBM83}
\begin{eqnarray}\label{eqB12}
   &&\sum\limits_{m = 0}^{\infty}\frac{t^m}{m!}I_{m + 1/2}(z) =  \Bigl
    ( \frac{2t}{z} + 1 \Bigr )^{- 1/4}I_{1/2}\Bigl (\sqrt {z^2 + 2tz }   \Bigr
    )\nonumber\\
    &&|z| - |2t| > 0\,.
\end{eqnarray}
In our case
\[|z| - |2t| = \frac{\delta |X|}{2\alpha(1 - \alpha ^2)}(\alpha ^2\gamma ^2 + 2 \alpha \gamma + \alpha ^2) > 0
\] and instead of equation (\ref{eqB11}) we have
\begin{equation}\label{eqB13}
{\mathscr J} = \sqrt{\frac{\pi}{\nu|\mu|}}\; e^{- \mu/2}I_{1/2}\Bigl
(\frac{\nu|\mu|}{2} \Bigr )\,,
\end{equation}
where
\begin{equation}\label{eqB14}
    \nu = \nu (\delta) =\sqrt{1 - \frac{\delta ^2}{\alpha ^2}(1 -
    \alpha ^2)}
    = \frac{\delta }{\alpha }\sqrt{\alpha ^2\gamma ^2 + 2\alpha\gamma
+ \alpha
    ^2}\,.
\end{equation}
Substituting the expression (\ref{eqB13})  into equation
(\ref{eqB6}) and taking into account that $I_{1/2}(x) = (2/\pi
x)^{1/2}\sinh (x)$, we obtain
\begin{eqnarray}
J^{(1)}_n &=& \frac{\delta _n}{\nu _n} \exp\Bigl [ \frac{X'}{2(1 -
\alpha)}\Bigr
    ]\times \nonumber\\
  & & \phantom{aaa}\times \Bigl\{\exp\Bigl [- \frac{\alpha (1 - \nu _n)X}{2\delta _n(1 - \alpha ^2)}{} \Bigr ] -
    \exp\Bigl [- \frac{\alpha (1 + \nu _n)X}{2\delta _n(1 - \alpha ^2)}{} \Bigr ]
    \Bigr\}\,,
 \label{eqB15}
\end{eqnarray}
 where $\nu _n = \nu(\delta _n)$. At last,
substituting $X'$ from Eq.~(\ref{eqB8}) into Eq.~(\ref{eqB15}) one
has
\begin{equation}\label{eqB16}
     J^{(1)}_n = \frac{\delta _n}{\nu _n} e^{- \gamma _n x/2}\Bigl\{ \exp\Bigl[\frac{\gamma_n^{(+)}X}{2(1 - \alpha^2)}\Bigr ]
     -
    \exp\Bigl[\frac{\gamma_n^{(-)}X}{2(1 - \alpha^2)}\Bigr ] \Bigr\}
\end{equation}
and \[\gamma_n^{(\pm)} =  \alpha + \gamma_n \pm
\sqrt{\alpha^2\gamma_n^2 + 2\alpha\gamma_n + \alpha^2}\,.\]

\subsection*{The calculation of $J_n^{(2)}$}
Consider the integral $J_n^{(2)}$. After substitution of the
variable in equation (\ref{eqB4})
\begin{equation}\label{eqB17}
    \xi + \frac{X}{1 - \alpha ^2} = x - x_1
\end{equation}
we have  (the index $n$ is omitted)
\begin{eqnarray}
 J^{(2)} &=& - \frac{X}{4\alpha}\exp{\Bigl [\frac{X'}{2(1 - \alpha)} - \frac{\alpha X}{2\delta (1 - \alpha ^2)} \Bigr ]}\times \nonumber\\
  &&\hspace{0.5cm}\times\int\limits_\frac{\alpha X}{1 - \alpha ^2}^{ - \frac{\alpha X}{1 - \alpha
    ^2}}d\xi   e^{\xi/2\delta } \frac{J'_0\Biggl (\frac{1}{2}
    \sqrt{{\displaystyle\frac{1 - \alpha ^2}{\alpha ^2}}}\sqrt{\Bigl ({\displaystyle\frac{\alpha X}{1 - \alpha ^2}\Bigr )^2} - \xi ^2} \; \Biggr )}
    {\sqrt{{\displaystyle\frac{1 - \alpha ^2}{\alpha ^2}}}\sqrt{\Bigl ({\displaystyle\frac{\alpha X}{1 - \alpha ^2}\Bigr )^2} - \xi^2}
    }\,.\label{eqB18}
\end{eqnarray}
To calculate  the integral (\ref{eqB18}) we consider the equality
(\ref{eqB15}), having previously  made the substitution
(\ref{eqB17}) into $J_n^{(1)}$. After reducing common factors, we
have
\begin{eqnarray}
& & \int\limits_{ \frac{\alpha X}{1 - \alpha ^2}}^{ - \frac{\alpha
X}{1 - \alpha
    ^2}}d\xi   e^{\xi/2\delta } J_0\Biggl (\frac{1}{2}
    \sqrt{{\displaystyle\frac{1 - \alpha ^2}{\alpha ^2}}}\sqrt{\Bigl ({\frac{\alpha X}{1 - \alpha ^2}\Bigr )^2} - \xi ^2} \; \Biggr
    ) = \nonumber\\
&&\hspace{6cm}= - \frac{4\delta}{\nu}   \sinh \Bigl [\frac{\alpha\nu
X}{2\delta(1 - \alpha ^2)}  \Bigr
 ]\,.\label{eqB19}
\end{eqnarray}
Differentiating the latter equation  with respect to $X$, one
obtains
\begin{eqnarray}
  \frac{X}{4\alpha}\int\limits_{ \frac{\alpha X}{1 - \alpha ^2}}^{ - \frac{\alpha X}{1 - \alpha
    ^2}}d\xi   e^{\xi/2\delta } \frac{J'_0\Biggl (\frac{1}{2}
  \sqrt{{\displaystyle\frac{1 - \alpha ^2}{\alpha ^2}}}\sqrt{\Bigl ({\displaystyle\frac{\alpha X}{1 - \alpha ^2}\Bigr )^2} - \xi ^2} \; \Biggr )}
    {\sqrt{{\displaystyle\frac{1 - \alpha ^2}{\alpha ^2}}}\sqrt{\Bigl ({\displaystyle\frac{\alpha X}{1 - \alpha ^2}\Bigr )^2} - \xi^2}
    }= \nonumber\\
  \phantom{aaaaaaaaaaaaaaaaa} = \cosh\frac{\alpha X}{2\delta (1 - \alpha ^2)}   -  \cosh\frac{\alpha\nu X}{2\delta (1 - \alpha ^2)}\,.\label{eqB20}
\end{eqnarray}
One multiplies the  latter equality  by
\[ - \exp\Bigl[ \frac{X'}{2(1 - \alpha)} - \frac{\alpha X}{2\delta (1 - \alpha ^2)} \Bigr] \]
and using equations (\ref{eqB18}), (\ref{eqB5}) and (\ref{eqB8}),
one has
\begin{eqnarray}
  J_n^{(2)} = \frac{1}{2}\,e^{-
\gamma_nx/2}\Bigl\{\exp\frac{\gamma_n^{(+)}X}{2(1 - \alpha ^2)} +
  \exp\frac{\gamma_n^{(-)}X}{2(1 - \alpha ^2)} \Bigr\} - \nonumber\\
   \hspace{1cm}- \frac{1}{2}\Bigl\{ \exp\Bigl [\frac{\alpha\gamma _n(x + t) + X}{2(1 - \alpha)}\Bigr ] +
   \exp\Bigl [- \frac{\alpha\gamma _n(x - t) + X}{2(1 + \alpha)}\Bigr ] \Bigr\}\,.\label{eqB21}
\end{eqnarray}
Finally, substitute equations (\ref{eqB16}) and (\ref{eqB21}) into
equation (\ref{eqB2})  and as a result we have

\begin{equation}\label{eqB22}
    \tilde{J}_n(x t) = e^{- \gamma_n x/2}\Bigl\{ A_n^{(-)}\exp\Bigl[\frac{\gamma_n^{(+)}X}{2(1 - \alpha^2)}\Bigr ] +
    A_n^{(+)}\exp\Bigl[\frac{\gamma_n^{(-)}X}{2(1 - \alpha^2)}\Bigr ]
    \Bigr\}\,,
\end{equation}
where
 \[ A_n^{(\pm)}= \frac{A_n}{2} \pm
 B_n\frac{\delta_n}{\nu_n}\,. \]

 \section{The determination of the parameters of the equations (\ref{eq36})-(\ref{eq39})}
For the  determination parameters entering (\ref{eq36})-(\ref{eq39})
we use  the condition continuity  of the temperature across the
interface, $\tilde{T}_{L} = \tilde{T}_{S}$, and  the condition of
the heat balance (\ref{eq12}).
 For this purpose initially we write down  the latent
heat of solidification $Q = kT_i$, with $T_i = T_L|_{X = 0} =
T_S|_{X = 0}$, in dimensionless form as
\begin{equation}
\label{eq40} \tilde{Q} = Q/Q_m = 1 + b\tilde{T}_S|_{X = 0}\, ,
\end{equation}
where  $b = Q_m/\rho c_p T_m$.

Now we equate the temperatures $\tilde{T}_S$ and $\tilde{T}_L$ at
the interface and substitute the fluxes (\ref{eq37}), (\ref{eq39})
at $X = 0$ into the condition of the heat balance (\ref{eq12}), then
taking into account the equalities (\ref{eq26}), one obtains
\begin{eqnarray}
&& ba_0^{(S)} + A_0/\alpha =  - 1\label{eq41}\\
&& a_0^{(S)} +  A_0/\alpha = \Delta \label{eq42}\\
&&A_n^{(S)} = \frac{n A_n}{n - b} \phantom{=.\frac{A_n}{n
- b}}\qquad (n \geqslant 1)\label{eq43}\\
&&\frac{A_{n + 1}^{(+)}}{n + 1} + \frac{A_{n}^{(+)}}{n} =
\frac{A_n}{n - b}\qquad (n \geqslant 1)\label{eq44}
\end{eqnarray}
From the last equation of (\ref{eq26}) and the equality (\ref{eq21})
it follows  that
\begin{equation}
 A_n^{(+)} - A_{n + 1}^{(+)} = A_n\,.
\label{eq27}
\end{equation}
 The substitution of this equality  into
(\ref{eq44}) gives the recurrent relationship
\[ A_{n + 1}^{(+)} =  \frac{b(n + 1)}{n(2n + 1 - b)}A_n^{(+)}\qquad (n \geqslant 1),\]
whence one obtains
\begin{equation}
A_n ^{(+)} =  - \frac{n b^{n - 1}}{(3 - b)(5 - b)\ldots(2n - 1 -
b)}A_0\qquad (n \geqslant 2),\label{eq45}
\end{equation}
where the equality $A_1^{(+)} = - A_0$ has been used (see the
relationships (\ref{eq26})). The remaining  parameters $A_0$ and
$a_0^{(S)}$ are found from the solution  of the system (\ref{eq41})
and (\ref{eq42}) in the form
\begin{equation}\label{eq46}
 A_0 = -  \alpha \frac{1 + b\Delta}{1 - b}\,,\qquad
 a_0^{(S)} =  \frac{1 + \Delta}{1 - b}\:.
\end{equation}
Finally, taking into account  the equalities (\ref{eq43}),
(\ref{eq45}) and (\ref{eq46}),   the expressions for the temperature
of both the liquid  and solid phase can be presented in the
dimensional coordinates $(x, t)$ in the form of the equations
(\ref{eq47}) and (\ref{eq48}).


\bibliographystyle{elsarticle-num}
\bibliography{<your-bib-database>}



\end{document}